\documentclass[12pt,preprint]{emulateapj}
\doublespace

\usepackage{epsfig}
\usepackage{natbib}
\usepackage{wasysym}
\usepackage{subfigure}
\usepackage{verbatim}
\usepackage{enumerate}
\usepackage{color}

\begin{document}

\title{ The Confined X-class Flares of Solar Active Region 2192}
\author{J.~K.~Thalmann\altaffilmark{1}, Y.~Su\altaffilmark{1}, M.~Temmer\altaffilmark{1} and A.~M.~Veronig\altaffilmark{1}}
\email{julia.thalmann@uni-graz.at}
\altaffiltext{1}{Institute of Physics/IGAM, University of Graz, Universit\"atsplatz 5/II, 8010 Graz, Austria}
\slugcomment{\today, accepted for publication in The Astrophysical Journal Letters}

\begin{abstract}
	The unusually large NOAA active region 2192, observed in October 2014, was outstanding in its productivity of major two-ribbon flares without coronal mass ejections. On a large scale, a predominantly north-south oriented magnetic system of arcade fields served as a strong, also lateral, confinement for a series of large two-ribbon flares originating from the core of the active region. The large initial separation of the flare ribbons, together with an almost absent growth in ribbon separation, suggests a confined reconnection site high up in the corona. Based on a detailed analysis of the confined X1.6 flare on October 22, we show how exceptional the flaring of this active region was. We provide evidence for repeated energy release, indicating that the same magnetic field structures were repeatedly involved in magnetic reconnection. We find that a large number of electrons was accelerated to non-thermal energies, revealing a steep power law spectrum, but that only a small fraction was accelerated to high energies. { The total non-thermal energy in electrons derived (on the order of $10^{25}$~J) is considerably higher than that in eruptive flares of class X1, and corresponds to about 10\% of the excess magnetic energy present in the active-region corona.}
\end{abstract}


\section{Introduction}

Coronal mass ejections (CMEs) and flares are interpreted to be different manifestations of a sudden instability and the associated release of magnetic energy in the solar corona. In general, they can occur independently of each other. Their association rate, however, is strongly increasing with the strength of the event. As can be inferred from Figure~1 of \cite{2006ApJ...650L.143Y}, in about 10\%, 40\%, and 75\% of {\it GOES} class C1-, M1-, and X1-flares, respectively, a CME association is found. Flares $\geq$X2.5 have an association rate $>$90\%. Sometimes, however, the Sun shows striking deviations from this trend.

On 2014 Oct~17, active region (AR) NOAA~2192 appeared on the east limb of the Sun and developed into the largest AR since NOAA~6368 in Nov 1990. In particular, the large size of NOAA~2192 was unexpected, as it occurred in unusually weak solar cycle 24. During its passage across the visible solar disk, between Oct~17 and 30, it produced six X- and 30 M-class flares, as well as numerous smaller events. The {\it GOES} soft X-ray (SXR) flux of the six largest flares peaked on Oct~19 05:03~UT (X1.1), Oct~22 14:28 UT (X1.6), Oct~24 22:41~UT (X3.1), Oct~25 17:08~UT (X1.0), Oct~26 10:56~UT (X2.0), and Oct~27 14:47~UT (X2.0). The highly exceptional aspect of the flaring activity was the lack of eruptive events: none of the X-flares was accompanied by a CME. 

In this letter, we investigate NOAA~2192 in the period Oct~22--24, regarding its productivity of a series of large ($\geq$M5) though confined flares and a single eruptive M4.0 flare. During this period, the AR was located within roughly $25^\circ$ from disk center so that foreshortening effects were minimal. Additionally, we analyze in detail the X1.6 flare on Oct~22 which, in contrast to the other X-flares during the considered period, was well covered also by hard X-ray (HXR) data.

\begin{figure*}
	\epsscale{1.0}
	\centering
	\includegraphics[width=2.0\columnwidth]{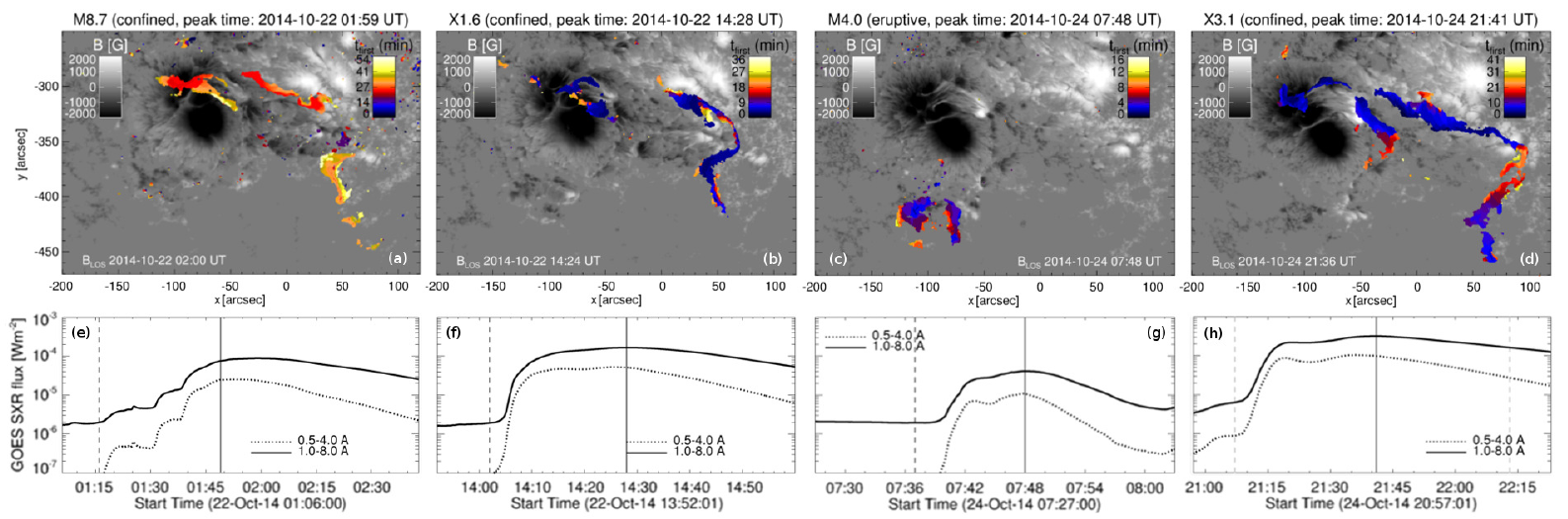}
	\caption{{\it Top panels:} Flare-ribbon progression associated to the (a) M8.7, (b) X1.6, (c) M4.0, and (d) X3.1 flare. The color indicates when a specific location was identified as a flare pixel for the first time in AIA 1700~\AA\ images (given in minutes after the flare onset). Locations are marked when the identification as flare pixel occurred between the start time and 10~min after the peak time of the respective flare. The gray-scale background resembles the LOS magnetic field of NOAA~2192 around the peak time of the respective flare, scaled to $\pm$\,2000~G. Black and white refers to negative and positive polarity, respectively. Units are arc-seconds from Sun center on Oct~23 12:00:00~UT. {\it Bottom panels:} {\it GOES} 0.5--4.0~\AA\ (gray dotted) and 1.0--8.0~\AA\ (black solid) SXR light curves of each flare. Vertical dashed and solid lines mark the start and peak time of the respective flare, respectively.}
	\label{f1}
\end{figure*}

\section{Data and methods}

We use data from the Atmospheric Imaging Assembly \citep[AIA;][]{2012SoPh..275...17L} on board the {\it Solar Dynamics Observatory} \citep[{\it SDO};][]{2012SoPh..275....3P}. In particular, 1700~\AA\ (sampling the photosphere at the temperature minimum) and 1600~\AA\ (picturing photospheric plus transition region emission) filtergrams were used for the analysis of flare ribbons. For the clear distinction of the low-atmosphere imprint of the four events under study, we use 1700~\AA\ data. In this way, we avoid a contamination of the signal due to ejected material during the eruptive M4.0 flare (which, in the line-of-sight, overlaps with the actual flare ribbons that we aim to track). For detailed analysis of the flare ribbons and the energy deposited by non-thermal electrons during the X1.6 flare, we use 1600~\AA\ images.

Short-term brightenings not related to flaring activity were removed by applying a 3-min running-median filter to the image sequences (at a 1-min cadence). These filtered images were used to track the location and time evolution of flare pixels. For the identification of flare pixels, we use the 99-percentile intensity of the entire series of filtered images as a threshold for detection. Importantly, the 99-percentile determines only the brightest pixels in a series of images in a particular wavelength due to its definition based on the relative occurrence of intensity values. Effects of blooming and saturated pixels around the flare peak time were minimized by demanding that a flare pixel has to be identified in at least five consecutive images.

The above data sets are complemented by {\it SDO}/Helioseismic and Magnetic Imager \citep[HMI;][]{2012SoPh..275..229S} magnetic field data. { The large-scale coronal magnetic field environment around NOAA~2192 is retrieved via the potential field source surface (PFSS) package available in {\it SolarSoftWare} \citep[for details see][]{2001ApJ...547..475S,2003SoPh..212..165S}. It is based on a synoptic HMI magnetogram for Carrington Rotation 2156 and gives the current-free coronal magnetic field between 1.0~$R_{\rm sun}$ and 2.5~$R_{\rm sun}$.} For the detailed analysis of the energetics involved in the X1.6 flare on Oct~22, the local coronal magnetic field in and around NOAA~2192 is approximated by a nonlinear force-free (NLFF) field, following \cite{2010A&A...516A.107W}.

This flare was also well covered by Reuven Ramaty High-Energy Solar Spectroscopic Imager \citep[{\it RHESSI};][]{2002SoPh..210....3L}. {\it RHESSI} X-ray images were reconstructed using the Clean algorithm \citep{2002SoPh..210...61H}. Additionally, we use AIA 94~\AA\ to trace the hot coronal flare plasma and ground-based H$\alpha$ filtergrams from Kanzelh\"ohe Observatory \citep[KSO;][]{2014arXiv1411.3896P} which sample purely chromospheric layers.

All data were prepared using standard IDL mapping software and corrected for the effect of differential rotation.

\section{Results}

\subsection{Flare ribbon progression -- confined vs.\ eruptive}

The bottom panels of Figure~\ref{f1} show the {\it GOES} SXR light curves for the investigated flares on Oct~22 and 24. All of the confined flares (Figure~\ref{f1}a, \ref{f1}b and \ref{f1}d) show gradual characteristics (an initial rise phase followed by a prolonged decay). The flare durations were significantly longer than that observed for the eruptive M4.0 flare (Figure~\ref{f1}c), in contrast to the previously reported impulsiveness of confined events \citep[e.~g.,][]{2006ApJ...650L.143Y,2011ApJ...732...87C}.

The top panels of Figure~\ref{f1} show the locations covered by flare ribbons, determined from 1700~\AA\ images. In the course of the confined flares under study (Figure~\ref{f1}a, \ref{f1}b and \ref{f1}d), two major ribbon systems are discernible: a shorter one close to the main negative-polarity sunspot and a longer one residing in the extended positive-polarity part of the AR.

The color code indicates when a certain position was identified as a flare pixel for the first time. Both ribbons appear first near the center of the AR and grow southward in time. This picture is clearly dominated by a large number of pixels brightening for the first time during the impulsive phase of the flares, when the ribbons grew fastest. However, despite showing a period of fast growth in extent, no considerable lateral separation of the ribbons was observed. Strikingly, they showed a large separation ($\gtrsim50$~Mm) already at the confined flares' onsets. For comparison, eruptive X-flares often show a ribbon separation of a few Mm in the rising phase, up to some tens of Mm in the decay phase \citep[e.\ g.,][]{2003ApJ...595.1251Z,2009SoPh..254..271X,2010SoPh..262..337M,2010ApJ...725..319Q}. We point out that the presented findings are neither a consequence of the wavelength selected for analysis, nor of the intensity threshold used to identify flaring pixels. 

Only the M4.0 flare had an associated CME and showed a clearly different location and morphology of the flare ribbons (Figure~\ref{f1}c). They were populating an area south to the main negative polarity, a region in which large-scale coronal loops seen in extreme ultraviolet (EUV) images (not shown here) fan out rapidly with their apexes reaching large heights and thus appear to be open. To understand why the major flares were confined, whereas the M4.0 flare was eruptive, we study the associated magnetic field topology.

\begin{figure}
	\centering
	\includegraphics[width=\columnwidth]{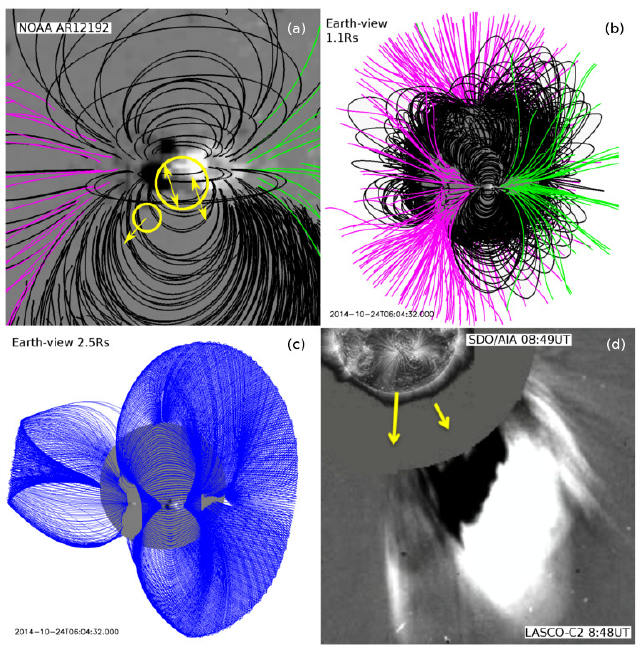}
	\caption{PFSS model result at 2014 Oct~24 06:04~UT. (a) Close-up view on the AR. Black lines indicate closed magnetic fields. Magenta and green lines mark open fields originating from locations of negative and positive polarity, respectively. The big yellow circle outlines the location of the high-energetic non-CME productive flares. The small yellow circle spots the location of the CME-productive flare. Arrows within these circles indicate the approximate direction of motion of the flare ribbons. (b) Model field lines using the starting point for field line calculation at 1.1~$R_{\rm sun}$. (c) Model field lines with starting point for field line calculation just below the source surface, i.~e., just below 2.5~$R_{\rm sun}$. The gray-scale background resembles the synoptic LOS magnetic field, scaled to $\pm$\,1200~G in (c) and to $\pm$\,600~G in (a) and (b). (d) {\it LASCO}-C2 image showing the CME associated to the M4.0 flare. An AIA 193~\AA\ image pictures hot flare plasma. Directions for the jet-like ejection and eruption are marked by a long and short yellow arrow, respectively.}
	\label{f2}
\end{figure}

\subsubsection{Large-scale magnetic field topology}

The magnetic field strength in sunspot umbrae usually ranges from 1000--2000~G, rarely exceeding 2500~G, with the umbral field strength scaling with the sunspot area \citep{2014SoPh..289.1477S}. NOAA~2192 well fits into this trend, given its umbral radius on the order of 10~Mm (as seen in AIA continuum images) and a line-of-sight (LOS) magnetic field magnitude of $\gtrsim$\,2600~G in the negative-polarity umbra. 

Figure~\ref{f2}a--\ref{f2}c show aspects of a PFSS model, timely centered on Oct~24 06:04 UT. In the close-up view in Figure~\ref{f2}a, we outline the location of the confined flares' ribbons (large yellow circle). Comparison to Figure~\ref{f2}b shows that they were situated below an arcade of strong field and all of them occurred in the AR core. This arcade, with apexes reaching up to the source surface (at 2.5~$R_{\rm sun}$; Figure~\ref{f2}c), likely prohibited the development of associated mass ejections. To substantiate, we calculated the potential field in and around NOAA~2192 up to a height of $\approx$1.5~$R_{\rm sun}$, using a fast-Fourier approach \citep{1981A&A...100..197A} based on HMI LOS magnetic field data. We calculate the total magnetic flux in a vertical plane, oriented along the main polarity inversion line. Following \cite{2007ApJ...665.1428W}, we employ the flux, normalized to the length of the vertical plane, in the two height regimes 1.0--1.1~$R_{\rm sun}$ ($F_{\rm low}$) and 1.1--1.5~$R_{\rm sun}$ ($F_{\rm high}$). The former measures the strength of the inner core field and the latter that of the overlying arcade field. We find $F_{\rm low}/F_{\rm high}\approx0.3$, indicating a strong constraint of the overlying field.

Repeating the calculation at the time and location of the eruptive M4.0 flare indicates a similarly strong constraint of the overlying magnetic field. In contrast to the confined flares, it occurred at the edge of the strong arcade fields, close to apparently open field structures (as seen in AIA 193~\AA\ image; compare Figure~\ref{f2}b and \ref{f2}d) towards which the flare ribbons progressed (Figure~\ref{f2}a). Indeed, AIA 193~\AA\ images (not shown here) reveal jet-like signatures, followed by a CME directed to the south-west of the AR (the direction of motion is indicated in Figure~\ref{f2}d). This favors a scenario in which the eruptive M4.0 flare was related to the interaction with neighboring open fields, rather than due to a weaker constraint by the overlying field.

 \begin{figure*}
	\epsscale{1.0}
	\centering
	\includegraphics[width=1.75\columnwidth]{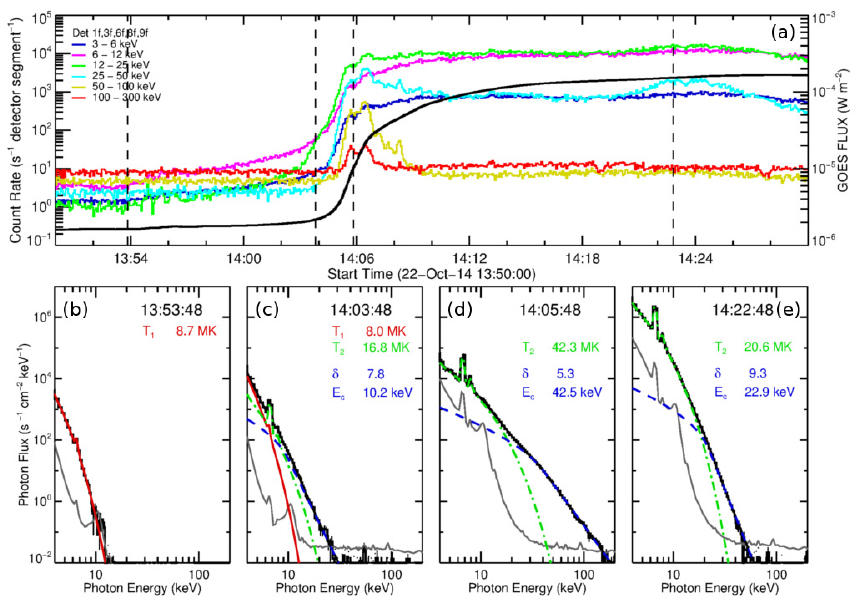}
	\caption{(a) {\it RHESSI} HXR count rates from 3~keV to 300~keV and {\it GOES} 1.0--8.0~\AA\ SXR flux (black solid line) during the X1.6 flare on Oct~22. Vertical dashed lines mark selected times for which X-ray spectra are shown: (b) before the flare onset, (c) during the rising phase, and around the two HXR peaks at $\sim$\,14:06~UT and $\sim$\,14:23~UT (panels (d) and (e), respectively). Panels (b)--(e) show the corresponding X-ray spectra (black solid lines) and fitting results. Gray solid lines represent the X-ray background. Red solid and green dash-dotted lines show fitted isothermal components. Blue dashed lines mark the fitted non-thermal component. The fitting parameters used, including temperature $T$, electron-distribution index $\delta$, and low energy cutoff $E_c$, are listed accordingly.}
	\label{f3}
\end{figure*}

\subsection{The confined X1.6 flare on Oct~22}

\subsubsection{Flare evolution}

Figure~\ref{f3}a shows the X-ray fluxes in the course of the flare. According to the {\it GOES} SXR light curve (black solid line), the impulsive phase of the flare started at 14:02~UT and the emission peaked at 14:28~UT. The {\it RHESSI} HXR $>$25~keV emission reveals two episodes of enhanced HXR bursts (around 14:06~UT and 14:23~UT). 

\begin{figure}
	\centering
	\includegraphics[width=\columnwidth]{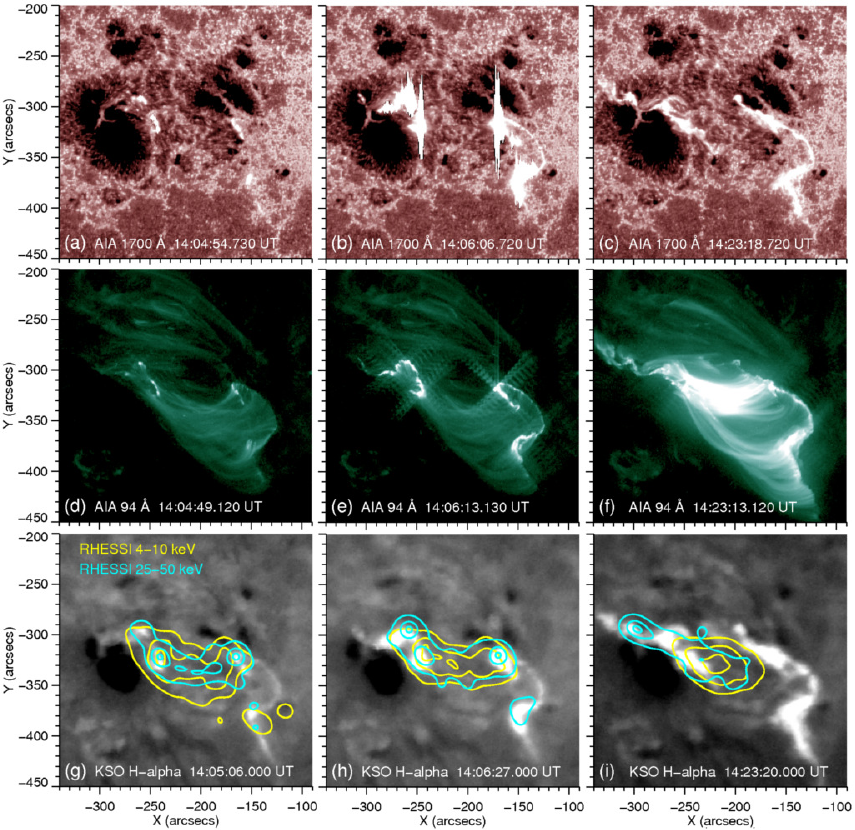}
	\caption{Coronal and low-atmosphere emission at three different times during the impulsive phase X1.6 flare on Oct~22. From top to bottom, ultraviolet (AIA 1700~\AA), EUV (AIA 94~\AA) and H$\alpha$ emission is shown. On top of the chromospheric H$\alpha$ images, in panels (g)--(i), {\it RHESSI} X-ray sources are shown. Yellow and cyan contours mark the emission in the 4--10~keV and 25--50~keV energy band, respectively, and are drawn at [10,50,90]\% of the respective maximum X-ray emission.}
	\label{f4}
\end{figure}

Figure~\ref{f4} shows the low-atmosphere and coronal emission during the impulsive phase of the flare. The AIA 1700~\AA\ (Figure~\ref{f4}a--\ref{f4}c) and KSO H$\alpha$ filtergrams (Figure~\ref{f4}g--\ref{f4}i) show the evolution of bright flare ribbons. AIA 94~\AA\ images (Figure~\ref{f4}d--\ref{f4}f) show a hot coronal flare loop system that connects these ribbons.

{\it RHESSI} X-ray images in the 4--10~keV and 25--50~keV energy bands (yellow and cyan contours, respectively) indicate the location of thermal and non-thermal sources, respectively (shown on top of H$\alpha$ images in Figure~\ref{f4}g--\ref{f4}i). Localized sources are seen already in the early phase of the flare (Figure~\ref{f4}g). The non-thermal sources are co-spatial with H$\alpha$ kernels, suggesting that these are footpoints of flaring loops, heated by non-thermal electron beams.

\subsubsection{Energetics}

To picture the evolution of the heated plasma and accelerated electrons, we show spatially integrated {\it RHESSI} X-ray spectra before the flare, during the rising phase and at the times of two HXR peaks in Figure~\ref{f3}b--\ref{f3}e. These spectra were fitted with an isothermal and a power-law non-thermal thick-target model \citep{2003ApJ...586..606H}. In the rise phase of the flare we used a second thermal component in order to achieve an acceptable goodness of the fit (Figure~\ref{f3}c). The non-thermal electron distribution is steep during the entire flare. The hardest spectrum occurred during the flare peak with an electron distribution index $\delta=5.3$ (Figure~\ref{f3}d). This means that a small number of the accelerated electrons reaches high energies and explains why the X-ray flux increase is limited to energies $<$300~keV.

Following \cite{2012ApJ...759...71E} and \cite{2013ApJ...765...37F}, we fit the {\it RHESSI} spectra (with a cadence of 20 seconds) between 14:03~UT and 14:34~UT, in order to estimate the non-thermal energies in flare-accelerated electrons. We find that the non-thermal electrons carried $\approx1.6\times10^{25}$~J. This is a factor of 10 larger than the energy in flare-accelerated electrons previously found for eruptive flares of {\it GOES} class X1 \citep[e.~g.,][]{2012ApJ...759...71E}. The uncertainty of such estimates in events with a large spectral index $\delta$, however, may be as large as one order of magnitude.

We compare the non-thermal energy estimate to the free magnetic energy stored in the AR. Assuming a force-free pre- and post-flare corona, we approximate the local corona of NOAA~2192 by a NLFF field. The magnetic energy of the NLFF field in excess over that of a corresponding potential field, gives an upper limit for the energy available for release during a flare. We consider a volume that covers the AR core (where the ribbons were observed) and extends up to $\approx90$~Mm, high enough to cover the reconnection site which is presumingly located somewhere below that height. { Prior to as well as after the X1.6 flare, we find an excess energy of $\approx1.5\times10^{26}$~J (with an estimated uncertainty of $\approx10\%$). This is in agreement with the high magnetic energies generally found for ARs hosting major flares \citep[see review by][]{2014A&ARv..22...78W}. Given the estimated non-thermal flare energy, roughly 10\% of the excess energy was carried away by accelerated electrons.} At the same time, however, magnetic energy was again stored and resulted in a similar amount of excess energy after the flare, allowing for equally intense energy releases during the following major events.

\begin{figure}
	\centering
	\includegraphics[width=\columnwidth]{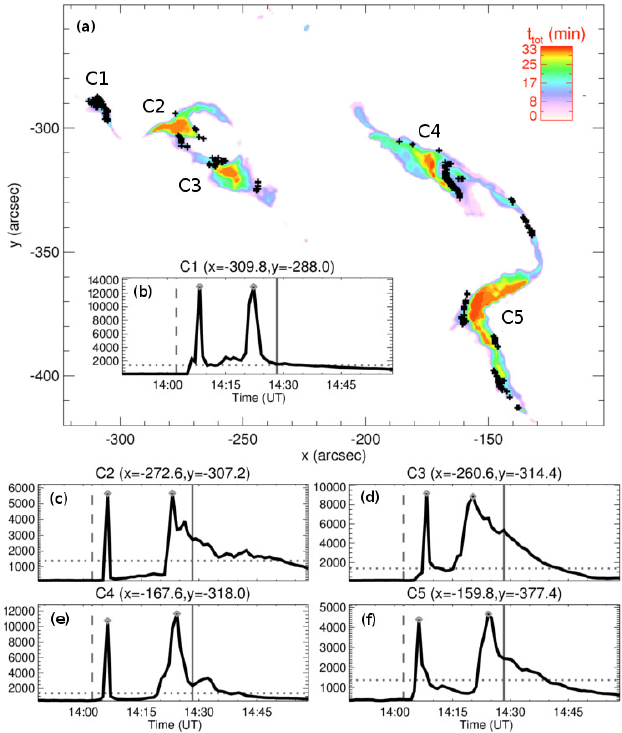}
	\caption{(a) Locations associated to flaring activity in the course of the X1.6 flare. Units are arc-seconds from Sun center on Oct~22 at 14:04~UT. The color code indicates how long individual locations showed enhanced emissivity, between the nominal start (14:02~UT) and end time (14:50~UT). Areas which were bright for the longest time are concentrated in five clusters (labeled $C_1$--$C_5$). Plus signs mark locations associated to repeated brightening. {Panels (b)--(f)} depict typical AIA 1600~\AA\ light curves at those locations. Vertical dashed and solid lines mark the start and peak time of the X1.6 flare, respectively. Horizontal dotted lines indicate the intensity threshold used to track flaring pixels.}
	\label{f5}
\end{figure}

\subsubsection{Recurrent brightening}

Figure~\ref{f5} shows the duration of brightness of the flare pixels tracked in AIA 1600~\AA. It appears that locations successively closer to the center of the ribbons were bright correspondingly longer. Longest flare emission, lasting up to $\approx30$~min, concentrated in five clusters (labeled as $C_1$--$C_5$). These locations nicely line up with the non-thermal {\it RHESSI} sources (cyan contours in Figure~\ref{f4}). 

Importantly, $C_1$--$C_5$ were associated to locations of recurrent brightenings, i.\ e., the re-energization of flare pixels (plus signs in Figure~\ref{f5}a). We interpret peaks detected in the 1600~\AA\ light curves as representing re-brightening whenever these peaks were separated in time by more than 7~min (to allow for cooling effects) and if more than one peak occurred at the same location during the impulsive phase (between 14:02~UT and 14:28~UT). In Figure~\ref{f5}b--\ref{f5}f, we show characteristic light curves of locations situated in clusters $C_1$--$C_5$, respectively. Inspection of the light curves reveals that the first intensity peak occurred in the early impulsive phase (before $\sim$\,14:10~UT), followed by another (sometimes even more pronounced) peak after $\sim$\,14:20~UT. Note that these periods of re-brightening are tightly associated to the two episodes of HXR bursts (Figure~\ref{f3}a). These findings evidence that magnetic field structures originating from a same narrow region (within the AIA resolution of $\approx$\,1$\farcs$2) were involved in multiple magnetic reconnection events. 

\section{Discussion}

NOAA~2192 showed an exceptional flaring behavior. In particular, it produced a series of six confined X-class flares in a period of nine days without associated CMEs. So far, only \cite{2007ApJ...665.1428W} reported five confined X-flares that originated from a single AR (in the course of two days). Using global magnetic field modeling, we find the cause of confinement in the form of a roughly north-south oriented arcade of strong magnetic field, serving as a top and lateral confinement to the flaring in the AR core. This is also supported by the more remote location of an eruptive M-class flare, which occurred close to the open field that neighbored the strong and closed core field. 

The flare ribbons observed during the confined major (M5.0 and larger) flares on Oct~22--24 exhibited a period of fast growth in extent but no considerable separation. This phenomenon was reported so far only for flares $<$M5.0 \citep{2007ApJ...655..606S}. In addition, the separation of the flare ribbons was large ($\approx$\,50~Mm) already at the flares' onsets, which suggests a reconnection site high in the corona. The associated SXR light curves classify the confined flares as long-duration events, objecting their suggested higher impulsiveness compared to eruptive events \citep[e.~g.,][]{2006ApJ...650L.143Y,2011ApJ...732...87C}.

Detailed analysis of the confined X1.6 flare on Oct~22 showed that the non-thermal electron distribution was very steep during the entire flare \citep[compare][]{2005A&A...439..737B} and that the total energy in electrons ($\approx10^{25}$~J) was, for an X1-flare, unusually high \citep[compare][]{2012ApJ...759...71E}. In accordance to previous studies, this pictures such events as efficient particle accelerators confined to the low corona \citep[e.\ g.,][]{2010SoPh..263..185K}. That also implies, however, that only a small fraction was accelerated to high energies, out of the large number of particles accelerated at the reconnection site. Comparison of the non-thermal flare energy and the magnetic excess energy in the AR shows that about 10\% of it was carried away by flare-accelerated electrons. Finally, we find re-brightening in flare pixels, providing evidence for the same magnetic field structures being repeatedly involved in magnetic reconnection.

\acknowledgments
\small
We thank the referee for careful consideration on this manuscript and useful comments. The authors acknowledge support from Austrian Science Fund (FWF): P25383-N27, P27292-N20 and V195-N16. JKT is thankful to J.\ M.\ Borrrero, J.\ Hirzberger, A.\ Lagg, and Y.\, Liu for helpful discussion on handling missing information in HMI data. {\it SDO} data are courtesy of the NASA/{\it SDO} AIA and HMI science teams. {\it RHESSI} is a NASA Small Explorer Mission. {\it GOES} is a joint effort of NASA and the National Oceanic and Atmospheric Administration (NOAA). H$\alpha$ data were provided by the Kanzelh\"ohe Observatory, University of Graz, Austria.

\end{document}